\documentclass[preprint,12pt]{elsarticle}

\usepackage{amssymb}

\usepackage{amsmath}
\PassOptionsToPackage{hidelinks}{hyperref}
\usepackage[nameinlink,noabbrev,english]{cleveref}
\usepackage{tikz,pgfplots,pgfplotstable} 
\usepackage{etoolbox}

\usepackage{siunitx} 
\usepackage{subcaption}
\PassOptionsToPackage{table}{xcolor}

\usepackage{colortbl}
\usepackage[skins,fitting,breakable]{tcolorbox}
\usepackage{xargs}                      
\usepackage{booktabs}

\usetikzlibrary{fit, calc,positioning,shadows,arrows,shapes,tikzmark,backgrounds}
\usetikzlibrary{external,decorations.pathreplacing,spy}
\pgfplotsset{compat=newest}
\pgfdeclarelayer{background}
\pgfdeclarelayer{foreground}
\pgfsetlayers{background, main, foreground}
\usepgfplotslibrary{colorbrewer, groupplots, fillbetween, statistics, units, polar}

\pgfkeys{/pgf/number format/.cd,precision=3, fixed, 1000 sep={}} 

\graphicspath{{./figures/latex}{./figures/images}}

\newcommand{\Uo}[1]{\qty{#1}{\metre\per\second}}
\newcommand{\Aventa}{\qty{7}{\kilo\watt} wind turbine}
\newcommand{\kindex}[2]{\ensuremath{{#1}_{\scalebox{0.5}{#2}}}}

\colorlet{ColorA}{GnBu-M}
\colorlet{ColorB}{GnBu-H}
\colorlet{ColorC}{GnBu-F}
\colorlet{ColorD}{GnBu-C}

\definecolor{cola1}{HTML}{00B0BE}
\definecolor{cola2}{HTML}{8FD7D7}
\definecolor{colb1}{HTML}{C99B28}
\definecolor{colb2}{HTML}{EDDCA5}

\definecolor{colt1}{RGB}{44,140,139} 
\definecolor{colt2}{RGB}{21,67,51} 
\definecolor{colt3}{RGB}{124,229,230} 
\definecolor{colg}{RGB}{242,163,38} 

\definecolor{AEROSENSEdarkblue}{RGB}{28,113,184}	
\definecolor{AEROSENSElightblue}{RGB}{28,113,184}
\definecolor{AEROSENSEblack}{RGB}{29,29,27}			
\definecolor{AEROSENSEgrey}{RGB}{198,198,198}		

\definecolor{OSThellgrau}{RGB}{198,198,198}
\definecolor{OSTschwarz}{RGB}{25,25,25}
\definecolor{OSTbrombeer}{RGB}{140,25,95}
\definecolor{OSThimbeer}{RGB}{215,40,100}
\definecolor{OSTdunkelviolett}{RGB}{107,56,129}
\definecolor{OSThellviolett}{RGB}{208,169,208}
\definecolor{OSTviolett}{RGB}{149,96,164}
\definecolor{OSTdunkelgrun}{RGB}{0,126,107}
\definecolor{OSThellgrun}{RGB}{167,213,194}
\definecolor{OSTgrun}{RGB}{29,175,142}
\definecolor{OSTdunkelrot}{RGB}{195,46,21}
\definecolor{OSThellrot}{RGB}{243,154,139}
\definecolor{OSTrot}{RGB}{232,78,15}
\definecolor{OSTdunkelblau}{RGB}{0,115,176}
\definecolor{OSThellblau}{RGB}{95,191,237}
\definecolor{OSTblau}{RGB}{0,134,205}
\definecolor{OSTdunkelorange}{RGB}{209,143,0}
\definecolor{OSThellorange}{RGB}{253,214,175}
\definecolor{OSTorange}{RGB}{251,186,0}

\journal{Journal of Wind Engineering and Industrial Aerodynamics}

\begin{document}

\begin{frontmatter}

    

\title{An aerodynamic measurement system to improve the efficiency of wind turbine rotor blades}

\author[1]{Julien Deparday}
\author[1,2]{Yuriy Marikovskiy}
\author[2]{Imad Abdallah}
\author[1]{Sarah Barber}
\affiliation[1]{organization={Institute for Energy Technology, OST - Eastern Switzerland University of Applied Sciences},
city={Rapperswil},
postcode={8645}, 
country={Switzerland}}
\affiliation[2]{organization={Chair of Structural Mechanics and Monitoring, Swiss Federal Institute of Technology (ETH Zurich)},
city={Zurich},
postcode={8093}, 
country={Switzerland}}


\begin{abstract}
    The wind energy sector is growing rapidly with the installation of wind turbines with long, slender blades in a diverse range of locations.
    To enhance the operational performance under specific wind conditions and to validate the aerodynamic design of flexible blades, it is crucial to obtain comprehensive data on the aerodynamic behaviour of the blades in the field, such as time-resolved pressure distributions, local inflow conditions and dynamic responses of the blade.
    However, published field measurements are scarce for large-scale rotor blades due to the complex and costly installation of the requisite measurement systems.
    In recent work, we developed a wireless and self-sufficient aerodynamic measurement system, named Aerosense, which is less complex and costly than conventional aerodynamic measurement systems.
    The Aerosense system uses Micro-Electro-Mechanical Systems (MEMS) sensors to obtain local aerodynamic pressures, blade motions, and inflow conditions.
    In this paper, we demonstrate the value of Aerosense in understanding the aerodynamic behaviour of rotor blades, using a \Aventa~operating in the field.
    After a thorough calibration and correction process, we demonstrate, for example, that the pressure distribution can vary significantly during one rotation of the blade, even under stable wind conditions.
    These variations are found to be due to the misalignment of the wind direction with the wind turbine's rotational axis.
    We therefore conclude that the Aerosense measurement system is valuable for understanding the aerodynamic loading on rotor blades as well as the influence of the inflow conditions on wind turbine performance.
\end{abstract}
    
    
    \end{frontmatter}

\section{Introduction}
\label{intro}

Wind energy is helping to reduce greenhouse gas emissions in the energy sector and is expected to continue to play a significant role in the future energy mix \cite{gwec2023}.
Wind turbines are increasing in efficiency and size, with rotor diameters now reaching up to \qty{200}{\meter}, and the rotor blades are becoming more flexible \cite{Veers2023GrandchallengesDesign}.
These large and slender rotor blades are subject to a wide range of spatial and turbulent scales in the wind, resulting in complex aerodynamic load fluctuations~\cite{Devinant2002, Schwarz_2019}.
The turbulent atmospheric boundary layer also results in a highly sheared flow, leading to further variations in wind speed and angle of attack along the rotor blade span, which yield varying spanwise aerodynamic loading on the blade.
Blade flexibility and wind variations cause cyclic loading, leading to fatigue and a reduction in the lifetime of rotor blades.
In addition, wind conditions (such as wind speed, shear and turbulence) are dependent on the geographical location and the surrounding topography.
It follows that site-dependent aerodynamic or control optimisation would be required to ensure optimal site-specific performance.
Local aerodynamic measurements in the field are therefore crucial to understanding the aerodynamic behaviour of the flexible blades for specific sites.

\subsection{Field measurements on rotating blades}
It has previously been shown that aerodynamic measurements in the field can improve the efficiency and the lifetime of wind turbines \cite{Schepers2019}.
The installation of aerodynamic measurement systems on rotating blades is a complex and expensive process due to the lack of access and space in the blades, as well as issues related to the wiring of sensors and data acquisition on a rotating system.
Therefore, published aerodynamic measurements on rotor blades in real conditions are very scarce, and no commercial measurement system is available for researchers, designers or owner/operators to carry out such measurements at a reasonable price.
To circumvent these issues, optical measurements installed on the ground that require no installation in the blade have been developed to measure the inflow conditions~\cite{Li2020} or the dynamic responses of the blade~\cite{Hoghooghi2020}.
However, the local aerodynamic loads on the rotating blades cannot be measured with such systems.

In order to measure local aerodynamic loading, the National Renewable Energy Laboratory (NREL) in the USA carried out extensive measurements on rotating wind turbine blades in the field in the 1990s \cite{hansen1993aerodynamics, simms1999unsteady}.
For example, they instrumented one blade with \num{60} to \num{64} pressure taps at nine different locations along the span to measure the local pressure distribution \cite{Medina2011}.
The pressure modules were housed in a temperature-controlled box that keeps a constant temperature for the sensors.
Four five-hole probes were installed at the leading edge to measure the local inflow angles and velocities.
Similarly, the Energy Research Centre of the Netherlands (ECN) conducted experimental campaigns on a \qty{300}{\kilo\watt} wind turbine in the 1990s.
They recently published a dataset of the acquired aerodynamic force and angles of attack at six different locations along the span of the blade \cite{ECN2023}. A group of universities have also conducted aerodynamic measurements during the Technology Collaboration Program (TCP) under the auspices of International Energy Agency (IEA)~\cite{task18report, Maeda2005SurfacePD}.
Surface pressure measurements and five-hole probes were also mounted and tested on an operating wind turbine in Japan \cite{Maeda2005SurfacePD}.
More recently, the Technical University of Denmark (DTU) conducted field experiments on a \qty{2}{\mega\watt} wind turbine \cite{Troldborg2013, Bertagnolio_semi-empirical_2017, schepers2021iea}. 
The blade was equipped with strain gauges, as well as with \num{64} surface pressure taps at four radial positions, with one five-hole probe close to each of these radial positions.
More recently, Wu et al. \cite{Wu2019} developed a field measurement system for a \qty{100}{\kilo\watt} wind turbine.
One of the blades was instrumented with flush-mounted pressure sensors at five different radial locations along the span, and seven-hole pitot tubes close to each of these locations.
Fritz et al. \cite{Fritz2024tiade} presented the long-term TIADE campaign on a \qty{3.8}{\mega\watt} wind turbine, on which 31 pressure taps were installed at 25\% of the blade radius.
A ground-based lidar and atmospheric measurements were used to infer the local inflow velocity.
All of these measurement systems required a dedicated instrumented blade, with pressure scanners inserted inside the blade, drilled holes, long tubings and cables running in the blade.
Such measurement systems are not suitable for industry applications, where the systems must be reasonably priced, easy to install and remove, and non-intrusive. 
As well as this, the pressure tubes limit the measurement frequency to approximately \qty{10}{\hertz} for \qty{25}{\meter} long tubes \cite{Medina2011}, which does not allow the unsteady effects to be examined.

\subsection{Correction and calibration on rotating blades}
The data processing for the existing systems mentioned above is also not straightforward.
Several calibration and correction procedures are required to obtain accurate aerodynamic measurements on rotating blades.
Medina et al.~\cite{Medina2011}, Schrek~\cite{Schreck2020} and Fritz et al.~\citep{Fritz2024tiade} describe the corrections necessary on a rotating blade for standard pressure scanners.
For example, the lengths of the tubings alter the pressure dynamics, the blade rotation yields a centrifugal force as well as a change of the hydrostatic pressure.
All of these parameters need to be taken into account and corrected.\\

The pressure distribution along the chord alone is not sufficient to obtain a comprehensive understanding of the aerodynamic performance.
Different inflow conditions may yield similar pressure distributions.
To be able to assess the causes of the aerodynamic performance, it is essential to infer the inflow conditions, such as the angle of attack and local wind speed.
However, obtaining local inflow conditions for rotating blades is challenging due to the complex nature of the local wind velocity at a given blade section.
The local velocity is a vector combination of the wind speed and the rotational speed of the airfoil, both of which are altered compared to the freestream conditions due to radial and axial induction through the rotor~\citep{Hau2013}.
To compare measurements, whether in the field or in wind tunnels, it is appropriate to use aerodynamic coefficients, which requires the knowledge of the dynamic pressure or local wind speed.
To compute the dynamic pressure, a common reference pressure must be defined.
The reference pressure is typically a static pressure measured at the hub, which is different to the atmospheric pressure due to axial induction~\cite{Medina2011,Fritz2024tiade}.
In the field measurements mentioned above, the blades were equipped with pitot tubes to measure the local inflow conditions.
Such systems are challenging to install, and the measurements are susceptible to induced velocity caused by bound circulation, and they need corrections for this upwash effect~\cite{everett1983seven, Wu2020}.\\

To summarise, previous measurement campaigns have demonstrated the potential value of aerodynamic measurements.
However, they also highlight the complexity and cost of embedding sensors inside a blade and retrieving the data via cables from a rotating machine.
They also demonstrate that the local pressure distribution must be measured together with the local inflow conditions to be able to analyse the measured aerodynamic pressure.\\

\subsection{Aerosense}
In this paper, therefore, we aim to demonstrate the value of a new measurement system, called Aerosense, which is easy to install, inexpensive and non-intrusive.
The hardware and installation make it a cost-effective solution for the wind energy industry. Previous work has demonstrated the functionality of the Aerosense hardware \cite{barber_aerosense_2022, Polonelli_AerosensePBL2023} and software \citep{marykovskiy2024architecting}; however, a detailed analysis of the measured data and a demonstration of the value of this data for rotor blade design has not yet been published.

The next section of this paper will present the measurement system and its characteristics.
Thorough calibration and correction procedures are required to obtain accurate aerodynamic measurements on rotating blades, which will be presented in the third section.
In the fourth section, we will then demonstrate the potential of this measurement system by presenting the experimental results conducted on a \Aventa.
It will also present a specific inference of the inflow conditions that does not require a reference pressure, and which allows finer analysis of the unsteady aerodynamic loadings.


\section{Measurement system}
\label{sec:ExpeSetup}
In this section, we first present an overview of the Aerosense system, followed by a description of the sensors.
It is also crucial to assess the blade shape at the location of the sensors in order to compute the integrated aerodynamic forces.
To this end, a method based on photogrammetry is also presented.

\subsection{General description of the Aerosense system}
The Aerosense measurement system (\cref{fig:Flowchart}) is based on low-cost, low-power MEMS (Micro-Electro-Mechanical Systems) sensors transmitting their data wirelessly to a base station \citep{Polonellietal2023AerosenseIMM}.
The advantages of this system are the easy installation, because of the absence of cables runninng along rotating blades, and the capacity to measure for a long period of time to obtain aerodynamic data for various wind conditions.
It also makes it possible to monitor the structural health of the blade \citep{barber_aerosense_2022}.
The system is composed of three main sub-systems:

\begin{figure}
    \centering    
    \includegraphics[width=0.95\textwidth]{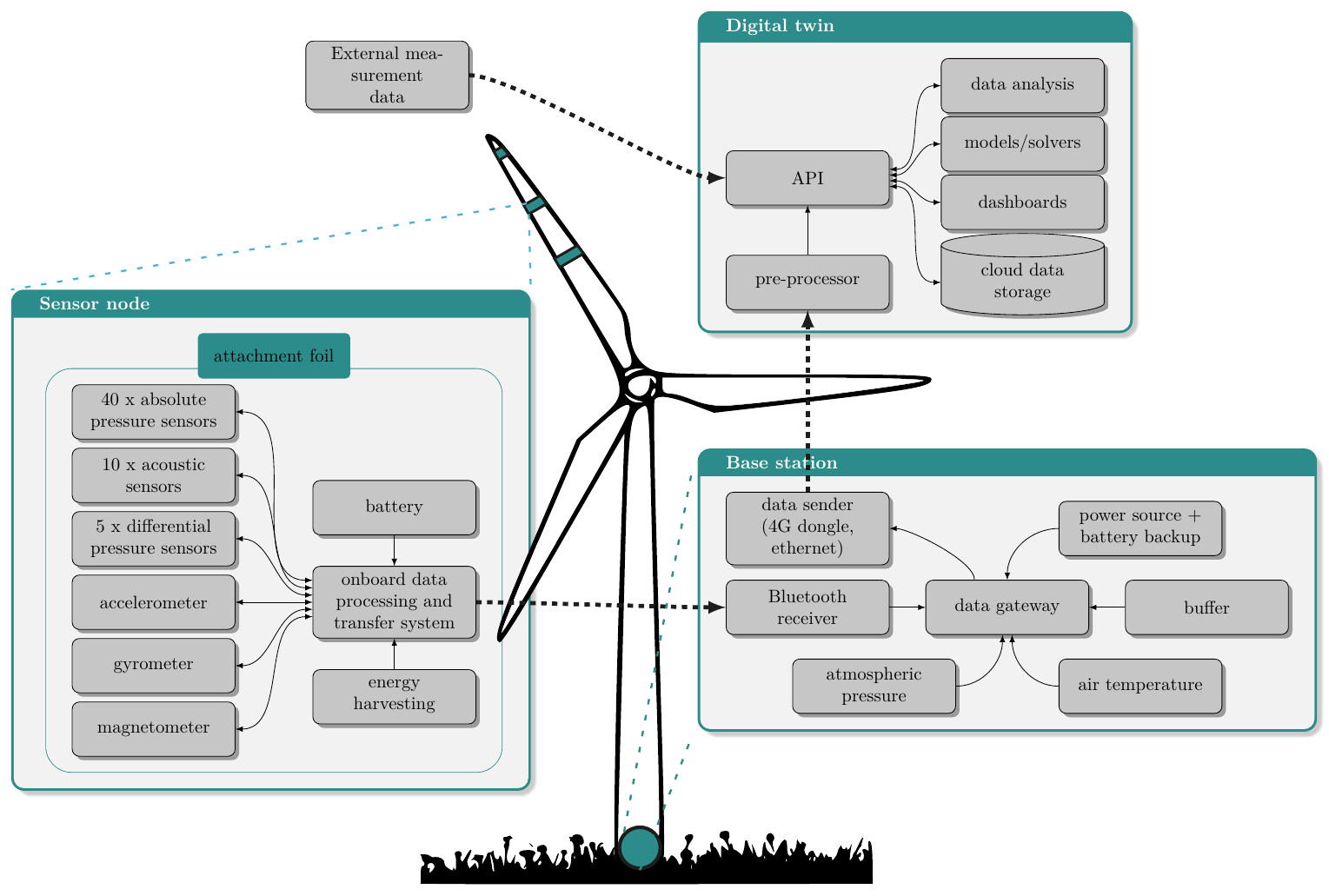}
    \caption{Flowchart of the Aerosense measurement system to measure aerodynamic properties on rotating wind turbine blades.}
    \label{fig:Flowchart}
\end{figure}

The first sub-system is the sensor node installed on the blades, which measures the local pressures and motions of the blades. 
Microphones for aeroacoustic purposes can also be installed near the trailing edge \citep{deparday_experimental_aeroacoustics_2022}, but will not be discussed in this paper.
All the sensors are connected to a microcontroller, which is the core of the system.
The microcontroller comprises a System on Chip (SoC) from Texas Instruments including three microprocessors, one as a main coordinator, one dedicated to support the digital sensors, and the last one dedicated to the Bluetooth Low Energy (BLE) protocol.
It is connected to an energy harvester system, which includes a \qty{5}{\mm} thick Lithium battery of \qty{1.85}{\watt\hour} and a \qty{360}{\milli\watt} flexible solar panel.
For more information on the electronic system, the reader is referred to \cite{Polonelli_AerosensePBL2023} and \cite{Polonellietal2023AerosenseIMM}.
The electronics are encapsulated in a thin flexible waterproof 3D-printed housing, which is easily installed on the blade using a leading-edge protection foil as tape integrated within the housing.
It has been designed to be thin (\qty{3}{\mm} maximum in the area where the aerodynamic pressure is measured, and \qty{5}{\mm} thick maximum at other locations) and therefore minimally aerodynamically intrusive.

The second sub-system is the base station, 
which can handle up to five sensor nodes with micro-second time synchronisation \citep{Polonellietal2023}.
It comprises a BLE transceiver and a local computing unit that collects the data from the sensor nodes and transmits it to a cloud storage.
The computational unit is based on a Raspberry Pi 4 with a Linux distribution using the Balena platform to be able to manage a fleet of base station devices and remotely interact with them over the web platform.
The base station also retrieves the atmospheric pressure and air temperature at the ground used as a reference.

The last sub-system is a digital twin system, which uses Octue SDK~\citep{octue_sdk} for integration of applications and orchestration of services, as well as Google BigQuery for data storage and querying.
It processes the field measurements, automatically analyses them, and can be used for the comparison and improvement of low and high-fidelity simulations.
It also provides a web-based visualisation of the data for quick access to data and results.
For more information on the software and pipeline, the reader is referred to Marykovksiy et al.~\cite{marykovskiy2024architecting}.

\subsection{Sensors}
\begin{figure}
    \centering  
    \def\svgwidth{0.5\linewidth}
\begingroup%
  \makeatletter%
  \providecommand\color[2][]{%
    \errmessage{(Inkscape) Color is used for the text in Inkscape, but the package 'color.sty' is not loaded}%
    \renewcommand\color[2][]{}%
  }%
  \providecommand\transparent[1]{%
    \errmessage{(Inkscape) Transparency is used (non-zero) for the text in Inkscape, but the package 'transparent.sty' is not loaded}%
    \renewcommand\transparent[1]{}%
  }%
  \providecommand\rotatebox[2]{#2}%
  \newcommand*\fsize{\dimexpr\f@size pt\relax}%
  \newcommand*\lineheight[1]{\fontsize{\fsize}{#1\fsize}\selectfont}%
  \ifx\svgwidth\undefined%
    \setlength{\unitlength}{172.18397366bp}%
    \ifx\svgscale\undefined%
      \relax%
    \else%
      \setlength{\unitlength}{\unitlength * \real{\svgscale}}%
    \fi%
  \else%
    \setlength{\unitlength}{\svgwidth}%
  \fi%
  \global\let\svgwidth\undefined%
  \global\let\svgscale\undefined%
  \makeatother%
  \begin{picture}(1,1.50042984)%
    \begin{footnotesize}
    \lineheight{1}%
    \setlength\tabcolsep{0pt}%
    \put(0,0){\includegraphics[width=\unitlength,page=1]{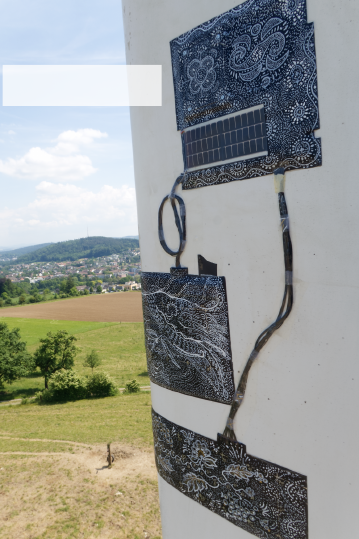}}%
    \put(0.01033576,1.28020745){\color[rgb]{0,0,0}\makebox(0,0)[lt]{\lineheight{1.25}\smash{\begin{tabular}[t]{l}data processing \\and transfer system\end{tabular}}}}%
    \put(0,0){\includegraphics[width=\unitlength,page=2]{FotoUkraineNode.pdf}}%
    \put(0.01115085,0.60360749){\color[rgb]{0,0,0}\makebox(0,0)[lt]{\lineheight{1.25}\smash{\begin{tabular}[t]{l}differential \\pressure sensors\end{tabular}}}}%
    \put(0,0){\includegraphics[width=\unitlength,page=3]{FotoUkraineNode.pdf}}%
    \put(0.00687435,0.23886605){\color[rgb]{0,0,0}\makebox(0,0)[lt]{\lineheight{1.25}\smash{\begin{tabular}[t]{l}absolute pressure \\sensors\end{tabular}}}}%
    \put(0,0){\includegraphics[width=\unitlength,page=4]{FotoUkraineNode.pdf}}%
    \put(0.2246447,1.06138395){\color[rgb]{0,0,0}\makebox(0,0)[lt]{\lineheight{1.25}\smash{\begin{tabular}[t]{l}solar panel\end{tabular}}}}%
    \end{footnotesize}
  \end{picture}%
\endgroup%

    \caption{Sensor node installed on a \Aventa.}
    \label{fig:AerosensePhoto}
\end{figure}

The pressure distribution and inflow conditions are obtained using various types of MEMS-based sensors installed on the blade as shown in \cref{fig:AerosensePhoto} and described in more detail below. 

\subsubsection{Absolute pressure sensors}
The pressure distribution on the blade is measured with an array of \num{40} absolute pressure sensors (ST LPS27HHW) installed along the chord on a chosen section of the blade.
The sensors from ST Microelectronics are water-resistant, only \qty{2.7}{\mm} thick and can measure from \qtyrange{260}{1260}{\hecto\pascal}.
They also possess an internal temperature calibration using an internal temperature sensor.
Absolute pressure sensors, also called barometers, measure the pressure relative to a vacuum chamber, which means that they do not need a reference pressure, unlike other pressure measurement systems installed in previous experimental campaigns \citep{Medina2011}. 
However, the disadvantage is that they measure not only the aerodynamic pressure, but also the hydrostatic pressure and the atmospheric pressure.
\Cref{sec:corrections} is dedicated to the correction of the pressure measurements and their post-processing to retrieve only the aerodynamic pressure.

According to the manufacturer, the absolute accuracy of the barometers is $\pm$\qty{50}{\pascal}, with a relative accuracy of $\pm$\SI{2.5}{\pascal}, which is barely enough to accurately measure the aerodynamic pressure on a wind turbine blade; the order of magnitude of the aerodynamic pressure on a multi megawatt wind turbine is about \SI{1000}{\pascal} (\cref{tab:PressuresBaros}), and therefore to achieve a 1\% accuracy, an absolute accuracy of \SI{10}{\pascal} should be reached for the full operating range of the wind turbine.
To enhance the accuracy of the sensors in all conditions, the barometers' pressure and temperature readings can be specifically calibrated to obtain a better accuracy of the pressure measurements.

The sampling frequency of the barometers is \qty{100}{\hertz}, which corresponds to about \num{500} measurement points for larger multi-megawatt wind turbines and to \num{100} measurement points during one rotation at the maximal rotational speed for the \Aventa, which is smaller and rotates faster.
This is sufficient to capture the main pressure fluctuations on a rotor blade.

\subsubsection{Differential pressure sensors}
Five differential pressure sensors measure differences in pressure at the leading edge of the blade.
These measurements are used as an input for a hybrid method inferring the inflow conditions \citep{Marykovskiyetal2023AerosenseAIAA}, based on the inviscid flow theory.
More details of this method are presented in \cref{sec:InflowCond}.
The sensors are from the 52-Series from Angst+Pfister and have a pressure range of \qtyrange{-2000}{2000}{\pascal} with a precision of $\pm$\qty{20}{\pascal}.
They also include an integrated multi-order compensation algorithm for correcting offset, sensitivity, and thermal errors.
They are \qty{5}{\mm} thick MEMS differential pressure sensors.
A sampling rate of \qty{1.2}{\kilo\hertz} is chosen to retrieve the dynamics of the turbulent inflow conditions.

\begin{figure}
    \centering  
    \def\svgwidth{0.5\linewidth}
    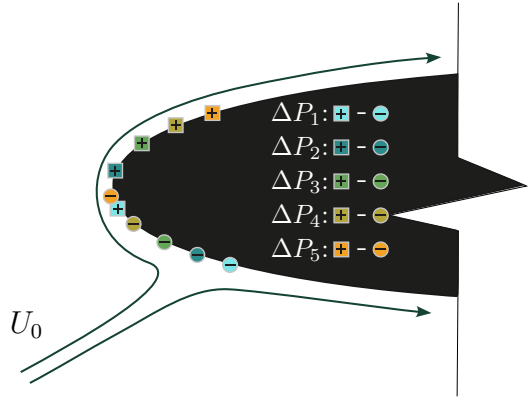
    \caption{Position of the differential pressure measurements around the leading edge.}
    \label{fig:Ideadifferential}
\end{figure}

This system measure pressures in the first 10\% of the chord (see~\cref{fig:Ideadifferential}) via short tubes connected to the sensors.
The sensors themselves are installed further downstream on the pressure side in order to minimise aerodynamic disturbances.
Each sensor is designed to measure a pressure difference between two ports.
The two ports of the sensor are connected via small tubes of \qty{2}{\milli\meter} outer diameter, which run within the flexible housing, to pressure taps that are small holes in the housing located at the leading edge.
For each sensor, the distance between the two taps is designed so that one hole is located on one side of the leading edge (for example the pressure side), and the other hole is positioned on the other side (the suction side).
The ideal curvilinear distance is approximately \num{10}\% of the chord length running around the leading edge of the airfoil.
For example in \cref{fig:Ideadifferential}, the first differential pressure sensor has its first hole on the pressure side at \num{10}\% of the chord, and the second hole, near the leading edge.
The other differential pressure sensors have their first hole distributed on the pressure side between \num{10}\% of the chord and the leading edge (see~\cref{fig:Ideadifferential}), with the same curvilinear distance between their two taps.

\subsubsection{Inertial measurement unit}
The last sensor type required to obtain accurate aerodynamic measurements on a rotating blade is a 9-axis inertial measurement unit (IMU) BMX160 from Bosch.
It consists of a 3-axis accelerometer, a 3-axis gyroscope, and a 3-axis magnetometer, with a sampling frequency of \qty{100}{\hertz}.
The measurements of the sensors are fused using a sensor fusion algorithm to obtain the blade motion and attitude \citep{Trummer2023IWASI}.
From the acceleration and rotational velocity measured at the blade section, it is possible to obtain the blade motion, the pitch angle and the azimuth position of the blade.
The azimuth position is used to correct the hydrostatic pressure measured by the barometers, and also to phase-average the pressure distribution for different azimuth angles, as explained in the next sections.

\section{Pressure corrections and geometry measurement}
\label{sec:corrections}

In this section, we describe the corrections applied on the raw data from the sensors to obtain aerodynamic pressure on a rotating blade.
In the next \cref{sec:PressureDist}, we present a real application of these corrections on the \Aventa.

\begin{table}
    \caption{Orders of magnitude of the types of pressure measured by absolute pressure sensors on a \qty{5}{\mega\watt} wind turbine.}
    \label{tab:PressuresBaros}%
    \resizebox{\textwidth}{!}{
    \begin{tabular}{lc|cc}
        \toprule
        Measured pressure & Symbol & Order of magnitude & Ratio with $P_{dyn}$  \\ 
        \midrule
        \rowcolor{colt1!25}
        Dynamic pressure & $P_{dyn}$ & \qty{2000}{\pascal} [200-7000] & 1 \\
        Atmospheric pressure & $P_{atm}$ & \qty{100000}{\pascal} & 50 \\
        Daily atmospheric pressure variations & $\Delta P_{atm}$ & \qty{2000}{\pascal} & 1 \\
        Pressure drift of barometers & $\Delta P_{drift}$ & \qty{100}{\pascal}/year & 0.05 \\
        Hydrostatic pressure variation & $\Delta P_{height}$ & \qty{800}{\pascal} & 0.4 \\
        Acceleration pressure variation & $K_{acc}$ & \qty{20}{\pascal} & 0.01 \\
    \end{tabular}
    }
\end{table}

To obtain valuable aerodynamic information from the sensors, a correction process must be applied.
The pressure measured by the barometers is the total pressure relative to a vacuum chamber located in the sensor ($P\eqsim0$).
The total pressure measured by the sensors $P_{\text{meas}}$ can be divided into the atmospheric pressure at hub height $P_{\text{atm}}$, the hydrostatic pressure due to the change of height when the blade rotates $\Delta P_{\text{height}}$, and the dynamic pressure due to the presence of the blade $P_{dyn}$. Local biases specific to the sensors (such as temperature) are generally taken into account in the sensor, but on rotor blades, sensors are used under specific conditions where they experience large accelerations, which may alter the dynamics of the inner membrane of the sensor by an amount $P_{acc}$:
\begin{equation}
     P_{\text{meas}}  = P_{\text{atm}} + \Delta P_{\text{height}} + P_{dyn} + P_{acc}.
 \end{equation}
The orders of magnitude of these different values are reported in \cref{tab:PressuresBaros}, based on estimations for a 5 MW wind turbine \cite{osti_947422},with a rated wind speed of \Uo{11}, a rotor diameter of \qty{126}{\meter} and a rotational speed of \qty{40}{rpm}.

\subsection{Correction for the atmospheric pressure and drift}
On a \qty{5}{\mega\watt} wind turbine, the atmospheric pressure is on the order of \num{50} times higher than the aerodynamic pressure.
The atmospheric pressure can be considered constant on a short measurement time (for example \qty{10}{\minute}), but may vary as much as the amplitude of the dynamic pressure in one day (if a cold front is coming for example) (\cref{tab:PressuresBaros}).
The atmospheric pressure is measured at the fixed base station and is used as a reference to evaluate the largest atmospheric pressure variations.

However, the barometers do not have the same reference (they all have their own vacuum chamber), therefore each sensor does not measure exactly the same atmospheric pressure, and their drift over time is usually not identical, as the vacuum chamber is never perfectly sealed.
Before installation, the sensors are therefore tested for a couple of weeks to estimate their individual drift. 
After installation of the sensor node on the blade, these corrections are verified when the blades are not rotating and with no wind.
The differences between every measured pressures in this standstill and no wind condition are then verified and readjusted if needed.

\subsection{Correction of the hydrostatic pressure}
The sensor rotates with the blade, which means the height above ground changes during the blade rotation, yielding a variation of the hydrostatic pressure, which is about 40\% of the aerodynamic pressure for a standard \qty{5}{\mega\watt} wind turbine.
The hydrostatic pressure is corrected based on a fusion algorithm using the measurements of the inertial measurement unit (IMU) installed at the same location as the absolute pressure sensors.
Specific fusion algorithms dedicated to wind turbine blades have to be developed \citep{Berkemeyer_imu_2021,Grundkötter_imu_2022,Plaza_imu_2023}, because unlike more standard fusion algorithms \citep{Valenti_imu_2016,Ghobadi_imu_2018} mostly dedicated to drone applications, wind turbine blades experience a large centrifugal acceleration, \num{10} to \num{20} times higher than the gravitational acceleration.
Such large centrifugal accelerations are not taken into account in standard fusion algorithms.
The acceleration at the blade section, $\ddot{\boldsymbol{x}}_{\text {sensor}}$ is 
\begin{equation}
\ddot{\boldsymbol{x}}_{\text {sensor}}=\boldsymbol{g}+\overbrace{\dot{\boldsymbol{\omega}} \times \boldsymbol{r}}^{\text {angular }}+\overbrace{\boldsymbol{\omega} \times(\boldsymbol{\omega} \times \boldsymbol{r})}^{\text {centrifugal}}
\end{equation}
where $\boldsymbol{g}$ is the gravitational acceleration, $\boldsymbol{\omega}$ is the rotational velocity of the blade, and $\boldsymbol{r}$ is the radial position of the sensor.
The centrifugal force is measured by the gyroscope.
The remaining acceleration is due to the blade rotation and dynamics as well as the gravity.
As the gravitational acceleration has always the same direction, pointing downwards, it is possible to estimate the orientation of the accelerometer when it is rotating, and then deduce the azimuth position of the blade.
The reader is referred to \cite{Trummer2023IWASI} for more details on the algorithm used within Aerosense.
From the azimuth position and the radial position of the sensor, we calculate the height of the sensor $\Delta h$ to compute the hydrostatic pressure:
\begin{equation}    
    \Delta P_{\text {height}} =  \rho g \Delta h.
\end{equation}

\subsection{Influence of acceleration}
\begin{figure}
    \centering  
    \def\svgwidth{0.5\linewidth}
\begingroup%
  \makeatletter%
  \providecommand\color[2][]{%
    \errmessage{(Inkscape) Color is used for the text in Inkscape, but the package 'color.sty' is not loaded}%
    \renewcommand\color[2][]{}%
  }%
  \providecommand\transparent[1]{%
    \errmessage{(Inkscape) Transparency is used (non-zero) for the text in Inkscape, but the package 'transparent.sty' is not loaded}%
    \renewcommand\transparent[1]{}%
  }%
  \providecommand\rotatebox[2]{#2}%
  \newcommand*\fsize{\dimexpr\f@size pt\relax}%
  \newcommand*\lineheight[1]{\fontsize{\fsize}{#1\fsize}\selectfont}%
  \ifx\svgwidth\undefined%
    \setlength{\unitlength}{431.25bp}%
    \ifx\svgscale\undefined%
      \relax%
    \else%
      \setlength{\unitlength}{\unitlength * \real{\svgscale}}%
    \fi%
  \else%
    \setlength{\unitlength}{\svgwidth}%
  \fi%
  \global\let\svgwidth\undefined%
  \global\let\svgscale\undefined%
  \makeatother%
  \begin{picture}(1,0.82608696)%
    \lineheight{1}%
    \setlength\tabcolsep{0pt}%
    \put(0,0){\includegraphics[width=\unitlength,page=1]{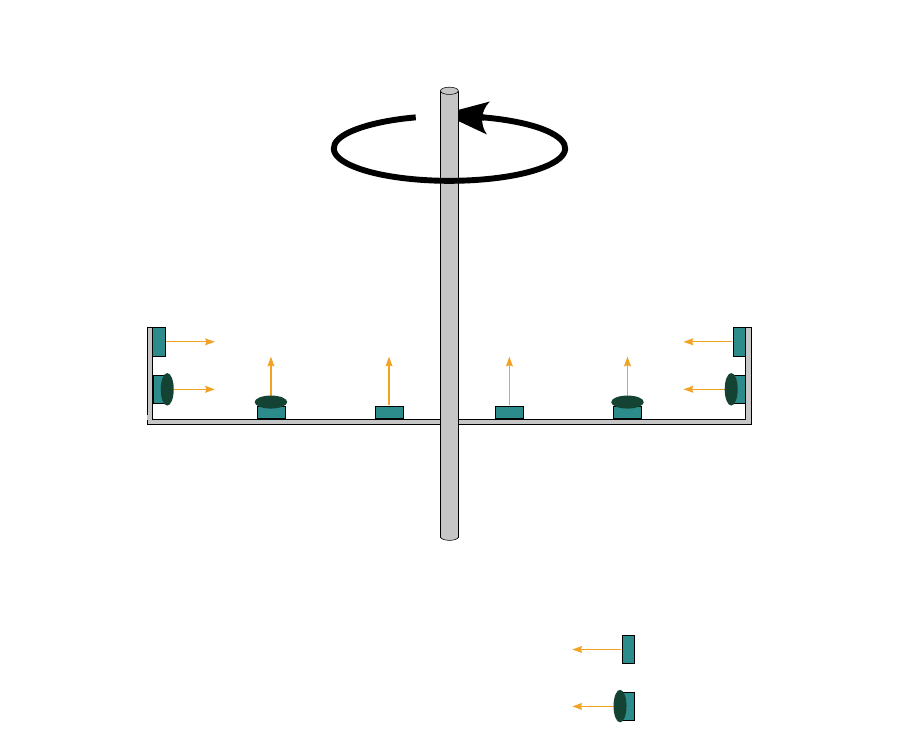}}%
    \put(0.2,0.0923412){\makebox(0,0)[lt]{\lineheight{1.25}\smash{\begin{tabular}[t]{l}\small opened sensor\end{tabular}}}}%
    \put(0.2,0.02931678){\makebox(0,0)[lt]{\lineheight{1.25}\smash{\begin{tabular}[t]{l}\small clogged sensor\end{tabular}}}}%
  \end{picture}%
\endgroup%

    \caption{Arrangement of the absolute pressure sensors around a rotational axis to assess the influence of the acceleration on readings. The orange arrow indicates the orientation of the sensors.}
    \label{fig:Sketch_acc}
\end{figure}

MEMS barometers are made up of a thin membrane that deforms in response to the pressure difference between the vacuum chamber and the external pressure.
The membrane's inertia may affect its deformation under strong acceleration.
Normally the influence of other parameters, such as temperature are taken into account during the conception and firmware of the MEMS barometers.
But, the use of such sensors experiencing large centrifugal force is not common and therefore not necessarily corrected.
Given the absence of any quantification of the influence of acceleration on barometer readings in the existing literature, we developed a small experimental setup to quantify the influence of centrifugal accelerations.
A centrifugal test bench was constructed and sensors were positioned in different orientations (\cref{fig:Sketch_acc}).
Some of the sensor holes were clogged by modelling clay to measure only the centrifugal acceleration.
The sensors oriented to the same direction of the rotational axis demonstrated no discernible influence of the centrifugal acceleration on their measurements.
This is the case of sensors installed on a blade of a horizontal axis wind turbine, where the centrifugal force points from root to the tip of the blade and is not normal to the sensor cavities.
Corrections of the centrifugal accelerations on pressure readings for horizontal axis wind turbine can therefore be neglected, and its correction is not taken into account for this study.

\begin{figure}
    \centering
    \includegraphics[width=0.7\textwidth]{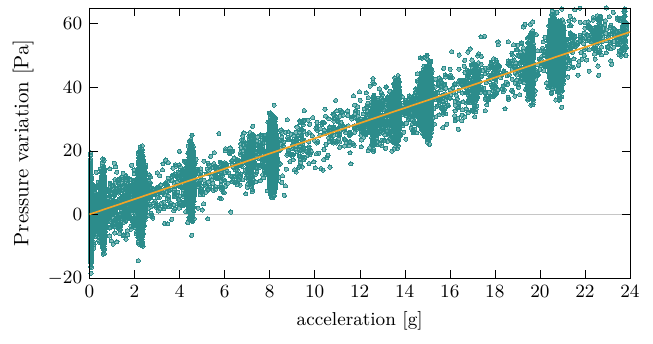}
    \caption{Influence of acceleration on the pressure measurements when the measurement cavity is in the direction of the acceleration. The slope of the linear fit (in orange) is \qty{2.4}{\pascal/g}.}
    \label{fig:BarosAccel}
\end{figure}

However, the sensors with the cavity hole facing the direction of the centrifugal acceleration exhibited a linear increase with the acceleration with a slope of \qty{2.4}{\pascal/g} (\cref{fig:BarosAccel}).
In the case of a vertical axis wind turbine, the influence of the acceleration should be taken into account, as the spanwise direction of the blades is in the same direction as the rotation axis.
Therefore, the sensors experience centrifugal accelerations during blade rotation.

\subsection{Blade geometry and sensor position measurement}
Eventually, in order to compute the pressure distribution and retrieve the aerodynamic force vector, the position of the sensors along a blade section is necessary.
The sensor node is manually installed on the blade, therefore the exact position of the sensors on the blade, as well as the geometry of the blade section, is not known.

We chose to use a photogrammetry process to measure the 3D shape and positions of the sensors.
Photogrammetry is often used to scan terrains using drones or airplanes \citep{AgueraVega_Survey_2017, Hagbo_windsimulations_2021}, but it can also be used to scan small objects and details \citep{Yravedra_Photogrammetry_2017,Gorski_ice_2020}, and large flexible objects \citep{deparday_full-scale_2016}.
With multiple photos taken from different angles, it is possible to triangulate the positions of the cameras as well as specific points placed on the blade.
A photogrammetry algorithm computes the position of the pressure sensors and analyses the precision of the measurement.
A section of the local 3D geometry of the blade is defined to determine the position of the pressure sensors along the chord of this section.


\section{Demonstration of the value of Aerosense measurements}
\label{sec:PressureDist}

The measurement system was installed and tested on a \Aventa~in Switzerland~\citep{Aventa2022and2023} between December 2022 and August 2023.
Some of the measurement data obtained during these tests is used in the present paper to compute the pressure distribution and inflow conditions on a rotor blade, and ultimately to demonstrate the potential of the Aerosense measurement system.
In this section, the corrections described in the previous section are applied on the barometer data, to then be able to compute the pressure distribution.
First, the corrections are applied to time-resolved absolute pressure data. 
Then the results are used together with the photogrammetry measurements to create phase-averaged pressure distributions. 
Finally, the inflow conditions are obtained to be able to analyse more finely the aerodynamic behaviour of the blade section.

\subsection{Calibration of the absolute pressure sensors}
In this project, the barometers were calibrated in a dedicated chamber at EMPA, the Swiss Federal Laboratories for Materials Science and Technology.
The calibration process involved a pressure range of \qtyrange{800}{1100}{h\pascal} and a temperature range of \qtyrange{-20}{50}{\celsius}.
As reference values, the pressure and temperature were measured with a precision of \qty{2}{\pascal} and \qty{0.1}{\celsius}.
It resulted in about \num{30000} measurement points of \qty{10}{\minute}, taken during \num{36} hours of continuous measurements.
A second-order surface fitting was applied to the measured temperature and pressure.
The Root Mean Squared Error between the reference pressure and the measured pressure value went from \qty{87}{\pascal} down to \qty{11}{\pascal}, which means that the pressure could be measured with a precision of 1\% for a multi-megawatt wind turbine.

\subsection{Application of the corrections to time-resolved absolute pressure data}
\label{sec:app_correction} 
\paragraph{Correction for the atmospheric pressure}

Before installation, the barometers were tested for several weeks in order to estimate their deviations.
Most of the sensors behaved with a linear drift over time with an average of \qty{0.77}{\pascal}/day. 
Four out of forty had a more erratic drift.
However, conditions with no wind were found and used as reference measurements, to recalibrate the reference atmospheric pressure.
The no wind condition was defined when the standard deviation of all the pressures is less than \qty{5}{\pascal} during \qty{10}{\min}, and the standard deviation for each axis of the accelerometer should be less than \qty{0.02}{\meter\per\square\second}.

\begin{figure}
    \centering
    \includegraphics[width=\textwidth]{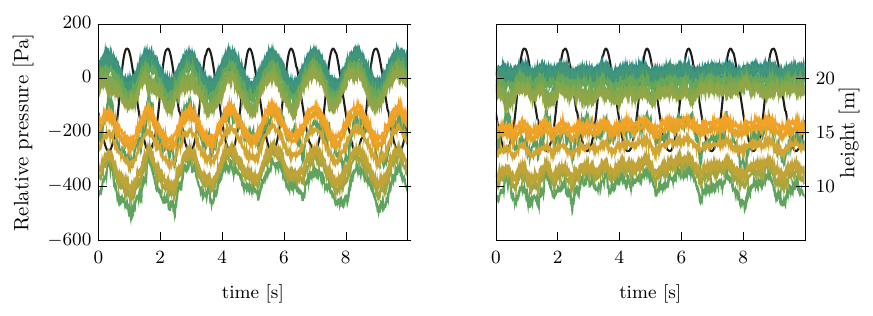}
    \caption{Pressure signals $P(t)$ at each measurement point after correction of the atmospheric pressure (left) and after the hydrostatic pressure (right) for the \Aventa. The black line in the background represents the height inferred by the IMU.}
    \label{fig:CorrectedPressure}
\end{figure}

\paragraph{Correction for the hydrostatic pressure}
During the measurement campaign, the fusion algorithm based on the IMU provided the height of the blade.
\Cref{fig:CorrectedPressure} shows the pressure signals $P(t)$ at each measurement point $x_i$ after correction of the atmospheric pressure (left) and after the hydrostatic pressure (right) using the equation:
\begin{equation}\label{eq:correctpressure}
    P_{\text{aero}}(x_i,t) = P_{\text {measured}}(x_i,t) - P_{\text {atm}}(x_i) + \Delta P_{height}(x_i,t)
\end{equation}
The height estimated by the IMU is shown with the black line in the background.
The main pressure fluctuations are in phase with the height signal.

\subsection{Measurement of the blade shape and sensor positions}
A GoPro camera filmed the complete installation on the \Aventa, from which \num{500} photographs were extracted for the photogrammetry process
As the blade is normally plain white, it was necessary to add texture to the blade by sticking white noise-printed images and finely decorating the sensor nodes.
These added texture and patterns greatly aided the photogrammetry reconstruction algorithm.
We used the software 3DFZephyr to obtain the 3D shape and analyse the precision of the measurement.

\begin{figure}
    \centering
    \begin{subfigure}[b]{0.4\textwidth}
        \includegraphics[width=\textwidth]{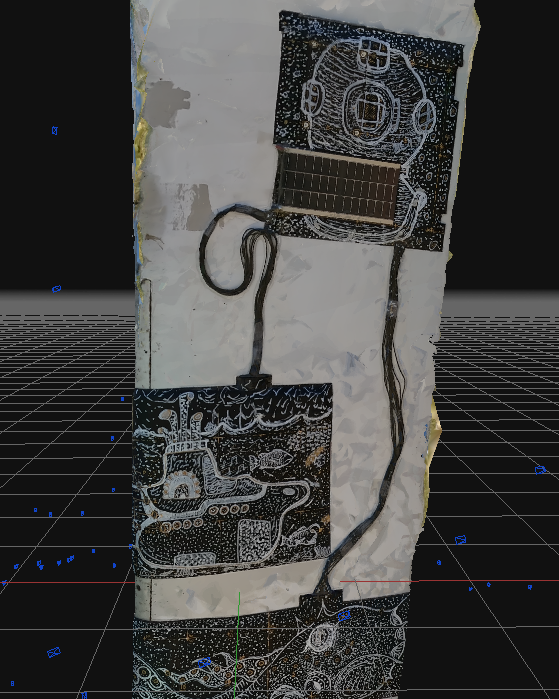}
        \caption{Global view of the 3D reconstruction of the Aerosense sensor node.}
        \label{fig:3Dglobalview}
    \end{subfigure}
    \hfill
    \begin{subfigure}[b]{0.4\textwidth}
        \includegraphics[width=\textwidth]{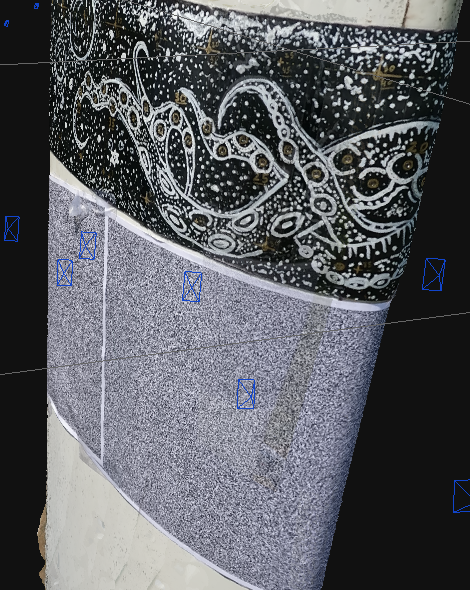}
        \caption{Close view of the 3D reconstruction of the absolute pressure sensors.}
        \label{fig:3Dcloseview}
    \end{subfigure}
    \hfill
    \begin{subfigure}[b]{0.6\textwidth}
    \includegraphics[width=\textwidth]{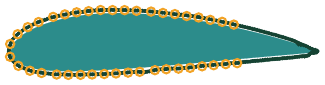}
    \caption{Section of the airfoil shape (theoretical shape: filled area, measured shape with sensor housing: dark green line) and the positions of the absolute pressure sensors (orange circles).}
    \label{fig:AirfoilShape}
    \end{subfigure}
    \caption{Photogrammetry reconstruction of the 3D shape of the blade and the positions of the sensors.}
    \label{fig:Photogrammetry}
\end{figure}

A 3D reconstruction of the blade is presented in \cref{fig:3Dglobalview} with a closer view of the absolute pressure sensors near the leading edge in \cref{fig:3Dcloseview}.
On the tested \Aventa, some edges and areas where the blade has no texture do not appear to be very well reconstructed (\cref{fig:3Dglobalview}), but the important areas such as the sensor nodes are correctly reconstructed with smooth surfaces (\cref{fig:3Dcloseview}). 
The position of the sensors is then precisely determined as well as the shape of the blade on that section (\cref{fig:AirfoilShape}).
The shape of the reconstructed blade (dark green line) is thicker than the theoretical shape (teal aerofoil), as it takes into account the thickness of the added housings onto the surface of the airfoil.
The photogrammetry algorithm was able to reconstruct the airfoil shape with an average relative error of \qty{0.76}{\mm} taking into account all the points.
To quantify an absolute accuracy, known distances, such as the distances between the sensors and other lengths measured on the \Aventa~were used to quantify the accuracy of the 3D reconstruction.
The absolute error ranged from \qtyrange{-1.15}{0.82}{\mm}, with an absolute average of \qty{0.14}{\mm}.
The chord length of the section is \qty{355}{\mm}, which means the blade shape and the positions of the sensors have an accuracy better than 0.5\% of the chord.

\subsection{Phase-averaged pressure distribution from field measurements}
\label{sec:res1}
\begin{figure}
    \centering
    \includegraphics[width=\textwidth]{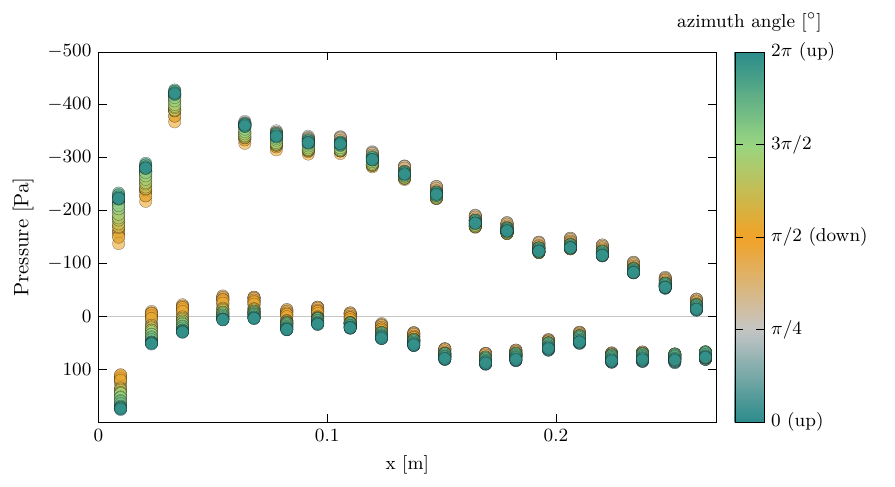}
    \caption{Phase-averaged pressure distribution along the chord of the blade for different azimuth angles.}
    \label{fig:PressureDistribution}
\end{figure}

The aerodynamic pressure is determined for every sensor, after the correction process.
Their position on the blade is also obtained thanks to the photogrammetry process.
In addition, the blade's azimuth position is obtained at the same frequency as the pressure signals, making it possible to compute the phase-averaged pressure distribution along the chord.
\Cref{fig:PressureDistribution} presents the phase-averaged pressure distribution for a stable period of \qty{1}{\min}, during which the local wind speed and blade rotation speed have a standard deviation of less than 5\% from their average.
The pressure distribution has the same features as a standard pressure distribution on a section of a wind turbine blade~\cite{Braudetal2024_PRF_StudyWallPressureVariations}.
A large suction area is found in the first 40\% of the chord. 
The aft half of the chord shows a slow decrease of the suction and tends to reach the same value as on the pressure side.
The pressure distribution appears to have a kink, for example on the pressure side at \qty{0.2}{\meter}, which does not make sense physically.
It corresponds to the position of the pressure sensors that had a more erratic drift than the others, as explained in \cref{sec:app_correction}.

The phase-averaged pressure distribution shows that the suction is at its maximum at the leading edge when the blade is pointing upwards (azimuth angle of \qty{0}{\radian}), whereas its minimum, with more than a 20\% decrease, occurs when the blade is pointing downwards (azimuth angle of \qty[parse-numbers = false]{\pi/2}{\radian}).
The pressure distribution varies much less with the azimuth angle in the aft half of the airfoil.
This will be discussed in the last section. 
But before the inflow conditions need to be estimated to be able to draw conclusions.

We demonstrated that the absolute pressure sensors can bring valuable time-resolved pressure distribution on a rotating blade after a thorough calibration and correction procedure, allowing a fine analysis of the aerodynamic loading on a rotor blade.
They show that, even for stable wind conditions, the phase-averaged pressure distribution can vary significantly with the azimuth angle, and the aerodynamic load is not constant during one rotation, which could cause premature fatigue of the blade and on the bearings.
However, the absolute pressure sensors provide only dimensionalised pressure values.
Without any information other than pressure distribution, it is impossible to understand the cause of these variations.

Non-dimensionalising the pressure distribution by the local wind speed and associating it with a local angle of attack would enable comparison with simulations or wind tunnel measurements on a fixed blade.
For example, if we want to compare with pressure measurements from a fixed wing in a wind tunnel experiments, the pressure is usually measured between an upstream reference pressure ($P_\infty$) and the pressure on the blade ($P_i$):
 \begin{equation}
     \Delta P_{ \text{meas,wt} } = P_i - P_\infty,
 \end{equation}
where the reference pressure is defined as follows
\begin{equation}
    P_{atm} + 0 =  P_\infty + q_\infty,
\end{equation}
where $q_\infty$ is the dynamic pressure.
Therefore, if we relate the pressure measurement in a wind tunnel with the atmospheric pressure:
\begin{equation}
    \Delta P_{ \text{meas,wt} } = P_{i} + q_\infty - P_{atm}.
\end{equation}

In our case on a rotating blade, the value $P_{atm}$ should also take into account the hydrostatic pressure (\cref{eq:correctpressure}).
To be able to correctly compare the pressure measurements with similar data in a wind tunnel or simulations, it is therefore required to determine the dynamic pressure $q_\infty$ which depends on the air density and the equivalent far-field wind speed $U_\infty$.

In the following section, we will describe a method for inferring the local angle of attack and wind speed on a rotating blade. 
We will also demonstrate how this additional information can enhance the insights gained from local aerodynamic measurements.

\subsection{Inflow conditions on rotating blades}
\label{sec:InflowCond}
The wind speed and the local angle of attack are complex to measure on a rotating blade, because the local inflow is the vectorial combination of the blade speed and wind speed, which progressively decreases when reaching the rotor plane.
This section presents a method to infer the local inflow conditions on a rotor blade using the differential pressure sensors.

\subsubsection{Overview of the hybrid method}
The incoming flow will hit the leading edge at the stagnation point with zero local velocity.
From the stagnation point on the pressure side, the flow goes around the leading edge and accelerates, yielding a large pressure gradient with a high suction on the suction side.
The stagnation point position and the pressure gradient depend on the shape of the leading edge, the angle of attack, and the wind speed.
The pressure gradient at the leading edge is measured with five differential pressure sensors.
Each of the sensors measures a difference of pressure between two positions at the leading edge of the blade (\cref{fig:Ideadifferential}).
For normal operating conditions, the flow is expected to be attached at the leading edge.
Assuming an inviscid attached flow at the leading edge, the pressure measured by the $i$th sensor can be written as:
 \begin{equation} \label{eq:Pdif}
     P_{i,1} - P_{i,2} = \frac{1}{2} \rho U_\infty^2 \left[ \left( \frac{U_2}{U_\infty} \right) ^2 - \left( \frac{U_1}{U_\infty} \right) ^2\right],
 \end{equation}
 where $P_{i,1}$, $P_{i,2}$ and $U_{i,1}$, $U_{i,2}$ are the pressure and the local velocity at the two measurement points of the $i$th sensor, respectively.
 $U_\infty$ is the far-field incoming wind speed, $\rho$ is the density of the fluid.

If the leading edge is approximated as a parabola, a conformal transformation can be applied to represent the flow around the leading edge as a flow near a flat plate.
The parabola defined in the space $z = x+iy$ is transformed as a flat plate in the space $\kappa = \epsilon + i \eta$.
In this new representation, the stagnation point position $\eta_S$ and the wind speed  $U_\infty$ can be analytically retrieved from the pressure measurements:
 \begin{equation} \label{eq:Pdif2}
    \Delta P_i = \frac{1}{2} \rho U_\infty^2 \left[ \frac{(\eta_2-\eta_S)^2}{1+\eta_2^2}  -  \frac{(\eta_1-\eta_S)^2}{1+\eta_1^2} \right].
\end{equation}
More detail on the method can be found in \cite{Marykovskiyetal2023AerosenseAIAA}, where it has been validated in a wind tunnel and its precision has been quantified within one degree when applied to an operating wind turbine in normal conditions.

This method has the advantage of not using any reference pressure $P_\infty$, which is more complicated to define and measure on a rotor blade than on a fixed wing.
It estimates the position of the stagnation point, which is a physical and well-defined property of the flow encountering any airfoil (the position on the blade where the local velocity is zero).
In contrast, the definition of the angle of attack on a rotor blade depends on the plane of reference, which is not well-defined due to the 3D rotating flow.
However, the equivalent angle of attack of a fixed 2D airfoil can be inferred from the stagnation point position using a look-up table based on wind tunnel measurements or simulations of a fixed airfoil.
This method allows a direct comparison of aerodynamic pressures from a rotating blade with a fixed 2D airfoil.

\subsubsection{Inflow condition inference on time-resolved field measurements}

\begin{figure}
    \centering
    \includegraphics[width=0.8\textwidth]{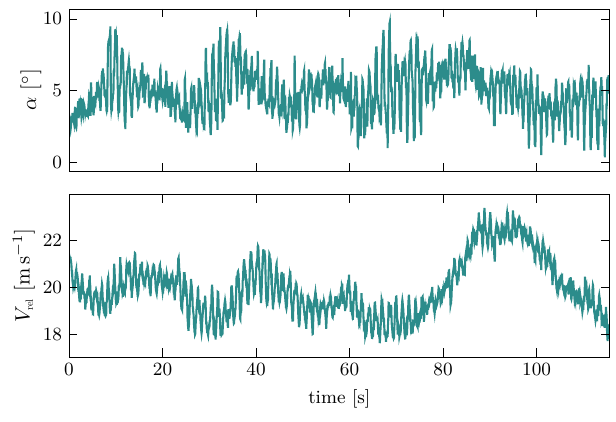}
    \caption{Evolution of the local angle of attack (top) and relative wind speed (bottom) encountered by the sensor on the rotor blade during two minutes.}
    \label{fig:diffPdata}
\end{figure}

The hybrid method has been applied to the field measurements on the \Aventa.
As an example, \cref{fig:diffPdata} presents the evolution of the inferred equivalent local angle of attack and the relative wind speed encountered by the sensor node on the rotor blade during two minutes of measurement.
The angle of attack varies mostly between \ang{3} and \ang{8}. 
A clear frequency is spotted, which represents the rotational speed of the blade.
The relative wind speed is about \Uo{20}.
The latest \qty{40}{\second} shows a wind gust with a momentary increase in wind speed of about 20\%.

\begin{figure}
    \centering
    \includegraphics[width=0.9\textwidth]{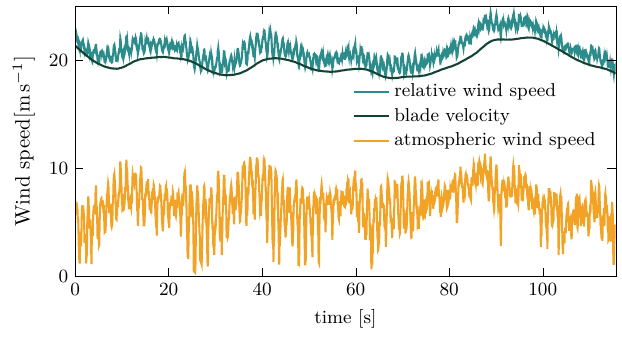}
    \caption{Estimation of the freestream wind speed by removing the local velocity of the blade from the relative wind speed.}
    \label{fig:IncomingWindSpeed}
\end{figure}    
Assuming a constant induction factor of 0.3 as previously found in the literature \citep{schepers2012analysis}, it is possible to estimate the atmospheric wind speed as presented in \cref{fig:IncomingWindSpeed}.
The relative wind speed is the vectorial combination of the blade velocity in dark green, defined as the rotational speed at the spanwise sensor location as well as local fluctuations due to the blade deformation, and the incoming atmospheric wind speed (orange curve).
The blade velocity estimated by the IMU is the main component of the relative wind speed, as the \Aventa~can turn up to \qty{1}{\hertz} yielding a large rotational speed.
The high frequency fluctuations are a consequence of the atmospheric wind speed.
It is important to note that the atmospheric wind speed is not defined in the reference frame of the wind turbine; rather, it is defined in the reference frame of the blade, which rotates.
The estimated atmospheric wind speed may differ as function of the azimuthal position of the blade due to, for example, yaw misalignment or shear flow.

\begin{figure}
    \centering
    \includegraphics[width=0.9\textwidth]{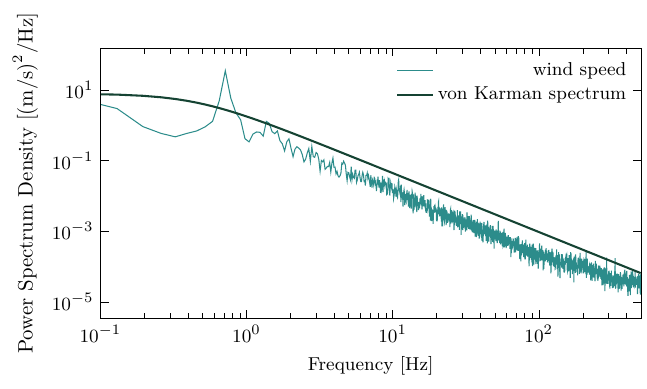}
    \caption{Power spectrum density of the estimated atmospheric wind speed.}
    \label{fig:PSD}
\end{figure}

The power spectrum density of the estimated atmospheric wind speed is presented in \cref{fig:PSD} for the same two minutes of measurement.
A broad frequency peak is found at \qty{0.7}{\hertz}, which corresponds to the rotational speed of the blade.
Between \qty{1}{\hertz} and \qty{500}{\hertz} the power spectrum density decreases linearly with a -5/3 slope, which is typical of a turbulent flow spectrum \cite{Morfiadakis1996}.
It is confirmed by the good fit of a von Kármán spectrum.
The von Kármán spectrum is generally suitable to represent turbulent wind for wind energy \citep{Acosta2024requirements}.
The von Kármán spectrum was built based on \citep{Otomo2024}:
\begin{equation}
    \phi(f)=\frac{4 u^{\prime 2}(L /\langle U\rangle)}{\left[1+70.8(f L /\langle U\rangle)^2\right]^{5 / 6}}.
\end{equation}
The von Kármán spectrum was generated from the statistics obtained from the wind measurements on the nacelle of the wind turbine. 
$u^{\prime}$ is the standard deviation of the wind speed, $\langle U\rangle$ is the mean wind speed ($\langle U\rangle$ = \qty{9.5}{\metre\per\second},$u^{\prime}/\langle U\rangle $ = 0.28), $L$ is the integral length scale of the turbulence ($L$ = \qty{2.5}{\m}).
The integral length scale was estimated based on Taylor’s frozen hypothesis as 
\begin{equation}
    L=\langle U\rangle \int_0^{\tau_L} \frac{\langle u(t) u(t+\tau)\rangle}{\left\langle u^2\right\rangle} \mathrm{d} \tau,
\end{equation}
where $\tau_L$ is the first zero-crossing location of the auto-correlation function.

The estimated atmospheric wind speed's power spectrum density appears to represent a realistic turbulent wind spectrum, suggesting that the inflow conditions inference method is capable of accurately estimating the incoming atmospheric wind speed and its turbulence.
Further validation with time-synchronised freestream wind measurements would be required to fully ascertain and validate it.

\subsubsection{Phase-average results}
\label{sec:res2}

\begin{figure}
    \centering
    \includegraphics[width=0.8\textwidth]{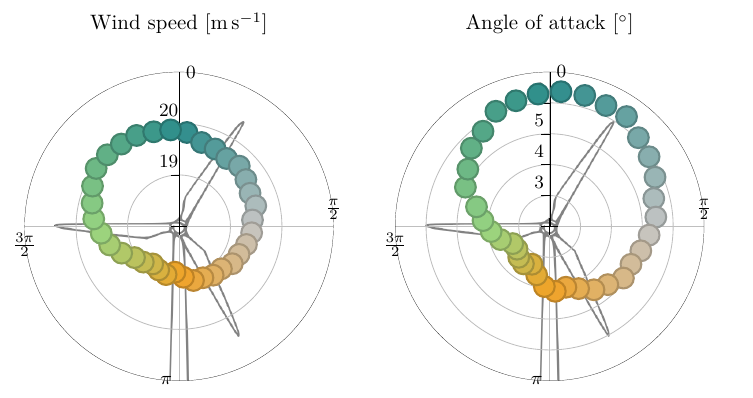}
    \caption{Polar representation of the local inflow conditions for different azimuth angles for the same stable conditions as presented in \cref{sec:res1}. Colours represent the azimuth position of the blade and eases the lecture of \cref{fig:PressureDistribution_xfoil}.}
    \label{fig:Polars}
\end{figure}

\begin{figure}
    \centering
    \includegraphics[width=\textwidth]{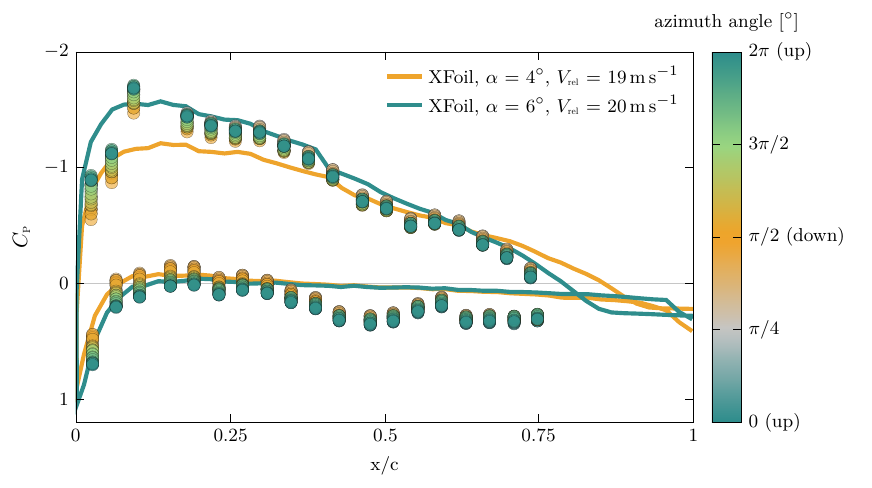}
    \caption{Phase-averaged distribution of the pressure coefficient for different azimuth angles with XFoil simulations on a fixed wing for two different inflow conditions.}
    \label{fig:PressureDistribution_xfoil}
\end{figure}

In the previous section, we showed that the incoming wind speed varies as function of the azimuth position of the blade and yields a distinct frequency peak.
We will now analyse the cause of this variation.
The time-resolved data allows us to analyse the local wind speed and angle of attack for various azimuth angles for the same stable conditions as presented in \cref{sec:res1}.
\Cref{fig:Polars} depicts the phase-average values of the relative wind speed (left) and local angle of attack (right).
At azimuth positions of 0 and \qty{2\pi}{\radian}, the blade is pointing upwards, and at \qty{\pi}{\radian}, the blade is pointing downwards.
The colours represent the azimuth position of the blade and helps in reading \cref{fig:PressureDistribution_xfoil}.

The wind speed and angle of attack distributions are not symmetrical between the first half when going down and the second half when going up.
The relative wind speed is maximum at \qty{20}{\metre\per\second} when the blade is pointing up and decreases to a minimum of \qty{19}{\metre\per\second} when the blade is pointing down.
The variation of wind speed of \qty{1}{\metre\per\second} between 0 and \qty{\pi}{\radian} is not large but may be due to the wind speed gradient within the atmospheric boundary layer (\cref{sec:AppA}).
A variation of \qty{1}{\metre\per\second} of the relative wind speed would require a variation of the atmospheric wind speed of about \qty{4}{\metre\per\second}, which is in the order of magnitude of the wind gradient.
However, the wind gradient within the atmospheric boundary layer is probably not the only reason, because the wind speed variation is not symmetrical between the first and second half of the rotation.

During most of the rotation, the angle of attack stays relatively constant between \ang{5} and \ang{6}.
But in the third quadrant ($\pi,3\pi/2$), the angle of attack decreases down to \ang{3} before reaching back \ang{6}, when the blade is pointing up.
Similar changes in the angle of attack during one rotation are also found in the literature when there is yaw misalignment, when the wind turbine is not facing the wind direction \cite{Wu2020, SotoValle2020}.
Dai et al.~\cite{Dai2017} showed that yaw misalignment causes a variation of the pressure distribution for different azimuth angles, with the main pressure variations located in the first 50\% of the chord, similar to the results presented in \cref{fig:PressureDistribution}.

Knowing the local wind speed and angle of attack, and also the blade profile, it is possible to non-dimensionalise the pressure distribution and compare it with simulations as presented in \cref{fig:PressureDistribution_xfoil}.
The local relative wind speed is used to compute the dynamic pressure.
It is also used as an input together with the angle of attack for XFOIl simulations.
XFOIL is able to compute efficiently pressure distribution on 2D airfoil using an inviscid method but which also takes into account viscous properties, such as transition from laminar to turbulent and limited trailing edge separations \citep{drela1989xfoil}.
XFOIL provides good results for attached flow \citep{morgado2016xfoil}, when the flow is still attached as it is for our case at an angle of attack of \ang{8} or lower.
Simulations are performed on the scanned geometry of the blade section for two different incoming wind speeds and angles of attack, representing the two extreme conditions encountered during the phase-average blade rotation. 
The simulation results are presented as solid lines in \cref{fig:PressureDistribution_xfoil}.
Overall, the simulations fits the experimental results.
The pressure gradients on the suction side are consistent between experiments and simulations.
The results diverge by almost half the dynamic pressure on the pressure side.
This might be mostly due to sensitivity and accuracy of the absolute pressure sensors that are not sufficient for such low dynamic pressure.
The amplitude of the pressure variations at the leading edge are well captured by the simulations and are in agreement with the results found in \cite{Dai2017}, when yaw misalignment occurs.
The simulations show that a small change of angle of attack (\ang{2}) and wind speed (\Uo{1}) may induce significant variations of suction at the leading edge.
The angle of attack variations is probably due to the yaw misalignment.\\

The Aerosense measurement system measures the pressure distribution and can also infer the inflow conditions in the field.
The combination of the two makes it possible to compare field measurements with simulations and allows for a finer analysis of complex flow dynamics and aerodynamic loading on a rotor blade.



\section{Conclusions}
\label{sec:Conclusions}

This paper introduces Aerosense, a wireless measurement system that acquires pressure distribution and local inflow conditions on rotating blades.
The MEMS-based measurement system is easy to install as it can be attached to the blade, without any modifications of the blade and any external wiring.
This makes it a low-cost system, valuable for acquiring more local aerodynamic data on rotating machines such as wind turbines to analyse their performance and therefore improve their efficiency and lifetime.

The measurement system includes absolute pressure sensors, differential pressure sensors, and an inertial measurement unit.
To obtain accurate aerodynamic pressure and inflow conditions, calibrations and correction processes are necessary.
After correcting the atmospheric and hydrostatic pressures, as well as using photogrammetry measurements to determine the position of the sensors and the blade shape, we demonstrate that we can obtain time-resolved pressure distributions.
These pressure measurements can be phase-averaged according to the azimuth angle inferred thanks to the in-house fusion algorithm for the inertial measurement unit.
Differential pressure sensors were utilised to infer the local inflow conditions, using a hybrid method based on the inviscid flow theory.
The method can estimate the atmospheric wind speed and its turbulence characteristics, which can aid in assessing the impact of inflow conditions on wind turbine performance.
The pressure distribution can be made non-dimensional by using the local dynamic pressure.
This allows for comparison with simulations or wind tunnel measurements on a fixed blade.

The Aerosense measurement system has been installed and validated on a \Aventa.
During the measurements on the wind turbine, even under stable wind conditions, the pressure distribution was found to vary significantly with the azimuth angle.
These variations could be explained by the non-alignment of the wind direction with the wind turbine's rotation axis.
This yaw misalignment caused a variation in the angle of attack during the blade's rotation, yielding load variations.

The Aerosense measurement system has therefore been shown to be a valuable tool for analysing the aerodynamic loading on a rotor blade and the influence of the inflow conditions on the wind turbine performance in the field.

\section{Acknowledgements}
The authors acknowledge the support from the Center for Project-Based Learning at ETH Zurich for the development of the electronics of the measurement system, and Stefan Rutzer and Josip Dubravac from OST Eastern Switzerland University of Applied Sciences, Institute for Materials Engineering and Plastics Processing for their help on the encapsulation of the sensors.

\section{CRediT author statement}
\textbf{Julien Deparday:} Conceptualization, Methodology, Software, Investigation, Writing - Original Draft, Writing - Review \& Editing, Visualization 
\textbf{Yuriy Marikovskiy:} Conceptualization, Software, Validation, Investigation, Data Curation, Writing - Review \& Editing 
\textbf{Imad Abdallah:} Validation, Investigation, Writing - Review \& Editing 
\textbf{Sarah Barber:} Conceptualization, Writing - Review \& Editing, Supervision, Project administration, Funding acquisition 

\section{Funding sources}
This work is funded by the BRIDGE Discovery Programme of the Swiss National Science Foundation and Innosuisse (Project Number
40B2-0\_187087).

\appendix 

\section{Estimation of the influence of the wind gradient on the relative wind velocity}
\label{sec:AppA}

Let's assume the relative wind speed (\kindex{V}{rel}) is \Uo{20} and the blade speed is \Uo{18} (\kindex{V}{bl}).
So we have an atmospheric wind speed ($V_{\infty}$) of \Uo{6} using the following equations and taking into account an induction factor of $\gamma = 0.3$ \cite{schepers2012analysis}:
\begin{equation}
    V_{\infty} = (1-\gamma)^2 \, \sqrt{\kindex{V}{rel}^2 - \kindex{V}{bl}^2}.
\end{equation}

If a change of relative wind speed, $d\kindex{v}{rel}$, is observed, the change of the atmospheric wind speed, $dV_{\infty}$, can be estimated as:
\begin{equation}
    (\kindex{V}{rel}+d\kindex{v}{rel})^2 = \frac{1}{1-\gamma} \left((V_{\infty}+dV_{\infty})^2 - \kindex{V}{bl}^2 \right).
\end{equation}

The relevant root of the polynomial yields:
\begin{equation}
    dV_{\infty} = V_{\infty} - \frac{1}{1-\gamma}\sqrt{(\kindex{V}{rel}+d\kindex{v}{rel})^2 - \kindex{V}{bl}^2}.
\end{equation}

For a change of relative wind speed of $d\kindex{v}{rel}$=\Uo{1}, the change of the atmospheric wind speed is about $dV_{\infty}$=\Uo{4}.

Let's now consider the wind gradient within the atmospheric boundary layer, and evaluate the difference of height expected for a change of wind speed of \Uo{4} at about the nacelle height (\kindex{z}{ref} = \qty{18}{\meter}).
The wind gradient is estimated using the power law \cite{Kikumoto2017observational}:
\begin{equation}
    V_{\infty}(z) = \kindex{V}{ref} \left(\frac{z}{\kindex{z}{ref}}\right)^{\alpha},
\end{equation}
where $\alpha$ is the power law exponent, which is about 0.3 for the atmospheric boundary layer, as estimated with local LIDAR measurements from previous experimental campaigns. 
A small change of the wind speed $dV_{\infty}$ will be due to a change of height $dh$:
\begin{equation}
    V_{\infty}(z_0) + dV_{\infty}  = V_{\infty}(z_0) \left(\frac{\kindex{z}{0} + dh}{\kindex{z}{0}}\right)^{\alpha},
\end{equation}

If $dh$ is small compared to $\kindex{z}{0}$, the equation can be linearised.
To only obtain an order of magnitude, the linearisation at the first order is sufficient:
\begin{equation}
    V_{\infty}(z_0) + dV_{\infty} = V_{\infty}(z_0) \left(1 + \alpha \frac{dh}{\kindex{z}{0}} + o(\frac{dh}{z_0}^2)\right).
\end{equation}
Therefore:
\begin{equation}
    dh \approxeq \frac{dV_{\infty}}{\alpha V_{\infty}(z_0)} \kindex{z}{0}.
\end{equation}
The change of height $dh$ is about \qty{11}{\meter} which corresponds to two times the radius where the sensor is placed on the blade (which has a radius of \qty{6}{\meter}).
Therefore, a change of the relative wind speed of \Uo{1} is in the order of magnitude of the wind gradient at the nacelle height for the maximum change of height experienced by the Aerosense sensor node during one rotation of the blade.

\bibliographystyle{elsarticle-num} 
\bibliography{biblio}

\end{document}